# Anomalous heat capacity and X-ray photoelectron spectroscopy of Superconducting $FeSe_{1/2}Te_{1/2}$

V.P.S. Awana[1], Govind[1], Anand Pal[1], Bhasker Gahtori[1], S.D. Kaushik[2], A. Vajpayee[1], Jagdish Kumar[1] and H. Kishan[1]

[1]Quantum Phenomenon and Applications Division, National Physical Laboratory (CSIR), Dr. K.S. Krishnan Marg, New Delhi 110012, India

[2]UGC-DAE Consortium for Scientific Research, Mumbai Centre, BARC, Trombay Mumbai – 400085, India

The bulk polycrystalline sample $FeSe_{1/2}Te_{1/2}$ is synthesized by solid state reaction route in an evacuated sealed quartz tube at 750 $^{o}$C. The presence of superconductivity is confirmed through magnetization/thermoelectric/resistivity studies. It is found that the superconducting transition temperature ($T_c$) is around 12 K. Heat capacity ($C_p$) of superconducting $FeSe_{1-x}Te_x$ exhibited a hump near $T_c$, instead of well defined Lambda transition. X-ray Photo electron spectroscopy (*XPS*) studies revealed well defined positions for divalent Fe, Se and Te but with sufficient hybridization of Fe (2p) and Se/Te (3d) core levels. In particular divalent Fe is shifted to higher BE (binding energy) and Se and Te to lower. The situation is similar to that as observed earlier for famous Cu based *HTSc* (High $T_c$ superconductors), where Cu (3d) orbital hybridizes with O (2p). We also found the satellite peak of Fe at 712.00 eV, which is attributed to charge carrier localization induced by Fe at 2c site.

Key Words: $FeTe_{1/2}Se_{1/2}$, Structure, Magnetization, Superconductivity, Heat Capacity, X-ray photo electron spectroscopy.
PACS: 74.25.-q, 74.62.Bf, 74.62.Dh

*Corresponding Author: Dr. V.P.S. Awana
Fax No. 0091-11-45609310: Phone No. 0091-11-45608329
e-mail-awana@mail.nplindia.ernet.in; www.freewebs.com/vpsawana/

## INTRODUCTION

Following the discovery of superconductivity in $LaFeAsO_{1-x}F_x$ ('1111') [1], an important development has been the invention of superconductivity in another class of materials Fe(Se,Te) ('11') [2]. Superconductivity has been observed in $FeSe_{1-x}$ system at ~8 K in its tetragonal form in the absence of doping [2]. Superconductivity in $FeSe_{1-x}$ system is significantly affected by applied pressure and chalcogenide substitutions [2-6]. In particular, an applied pressure of only up to 4.15 GPa enhances its $T_c$ to ~37 K with ($dT_c/dP$) of around 9K/GPa [6]. The effect of chemical pressure has been studied in '11' systems by means of Se-site substitution [4]. It is found that superconducting transition temperature increases with Te doping in $FeSe_{1-x}Te_x$ system, reaches a maximum at about 50% substitution, and then decreases with more Te doping [4]. Interestingly, FeTe is no longer superconducting. Here we report the synthesis, structure, magnetization, heat capacity and X-ray photo electron spectroscopy (*XPS*) studies of $FeSe_{1/2}Te_{1/2}$ compound, which is the '11' compound with the highest $T_c$ value of ~14.5 K.

## EXPERIMENTAL

The polycrystalline sample of $FeSe_{1/2}Te_{1/2}$ was synthesized by the solid state reaction route. The stoichiometric ratio of highly pure (> 3N) Fe, Se, and Te are ground, pelletized and then encapsulated in an evacuated ($10^{-3}$ Torr) quartz tube. The encapsulated tube is then heated at 750 $^oC$ for 12 hours and slowly cooled to room temperature. The initially sintered powder was again ground, pelletized in a rectangular shape and sealed in an evacuated quartz tube and re-sintered at 750 $^oC$ for 12 hours. The x-ray diffraction patterns of the samples are obtained with the help of a Rigaku diffractometer using $CuK_\alpha$ radiation. The resistivity measurements are recorded for temperatures down to 4.2 K via four probe method. Temperature dependence of AC magnetization of the $FeSe_{1/2}Te_{1/2}$ sample is taken on *PPMS* (physical property measurement system). Heat capacity $C_P(T)$ in zero field is also measured on PPMS. The $FeSe_{1/2}Te_{1/2}$ has been characterized by x-ray photoelectron spectroscopy (*XPS*) Perkin Elmer -PHI Model 1257, working at a base pressure of 5x $10^{-10}$ torr. The chamber is equipped with a dual anode Mg $K_{\alpha}$(1253.6 eV) and Al $K_{\alpha}$(1486.6 eV) x-ray sources and a high-resolution hemispherical electron energy analyzer. We have used Mg $K_{\alpha}$ x-ray source for our analysis. The calibration of the binding energy scale has done with the C (1s) line at 284.6eV. The core level spectra of Fe, Te and Se have been de-convoluted in to the Gaussian components.

## RESULTS AND DISCUSSION

Figure 1 depicts the observed and fitted X-ray diffraction pattern of $FeSe_{1/2}Te_{1/2}$ that corresponds to P4/nmm space group. The Rietveld refinement was performed using FULLPROF SUITE program and structure was fitted in P4/nmm space group with lattice parameters a = 3.7926(3) Å and c = 6.015(2) Å. These values are in agreement with earlier reports [7, 8]. The refinement was done taking Fe ions at two different sites (2a and 2c). The occupation of Fe at interstitial 2c site was found to be about 8% [9]. The insets of Figure 1 show the resistivity (ρ) and AC susceptibility (χ) versus temperature (*T*) plots for the studied $FeSe_{1/2}Te_{1/2}$ compound. Both ρ(*T*) and χ(*T*) demonstrate that the compound is bulk superconducting below 12 K. Detailed physical property characterization of the studied $FeSe_{1/2}Te_{1/2}$ compound is reported by some of us previously [10].

Figure 2 shows the heat capacity ($C_P$) versus temperature (*T*) plot for $FeSe_{1/2}Te_{1/2}$. The room temperature (300 K) $C_P$ is around 55 J/mol-K, which is in general agreement with reported values for this compound [9]. The data was fitted to the equation $C_p = \gamma T + BT^3 + CT^5$ from above superconducting transition from 13 K to 16 K. Where, $\gamma T$ and $BT^3+CT^5$ are electronic and phononic specific heat contributions respectively. The value of γ was found to be 57.73 mJ/mol-$K^2$ which is somewhat higher than other reported values [11-12]. However, the value of Debye temperature (171 K) is in agreement with their reported values [11]. From this fitting we estimated corresponding normal state behavior below $T_c$. The lower inset shows measured value of $C_p/T$ vs. $T$ and corresponding normal state fitted value extrapolated up to 2 K. As far as the entropy contribution due to superconducting condensate is concerned, the superconducting state – normal state is plotted in upper inset of Figure 2. At superconducting transition temperature ($T_c$) the discontinuity in electronic $C_p$ vs. $T$ plot is seen and marked. The shape of Lambda transition for studied $FeSe_{1/2}Te_{1/2}$ does not exhibit sharp discontinuity at $T_c$ as seen for other superconductors [13, 14], rather a broad hump like structure [13].

X-ray photoelectron spectroscopy studies, revealed well defined positions for Fe, Se and Te but with sufficient hybridization. Figure 3 (a), 3 (b) and 3 (c) shows the deconvoluted core level *XPS* spectra of Fe (2p), Te (3d) and Se (3d) respectively for $FeSe_{1/2}Te_{1/2}$ sample. The Fe ($2p_{3/2}$) spectra (Fig 3a) could be resolved into three spin-orbit components at 706.65 eV, 708.50 eV and 712.00 eV after the carbon correction. A Fe (2p1/2) core level spectrum was also deconvoluted three peaks in similar way. The 3d core level spectrum of Te is deconvoluted into three peaks with binding energy 572.00 eV, 572.65 eV and 573.90 eV. The peak at energy 572.65 eV corresponds to pure Te metal while peak at lower binding energy 572.00 eV may have appeared due to the hybridization. In case of Se, the core level spectra are deconvoluted into two peaks with binding energies 53.52 eV and 55.28 eV. The peak for higher BE corresponds to pure Se while peak at lower BE appeared due to the hybridization.

In Figure 3 (a) the peak at binding energy 706.65 eV corresponds to pure Fe metal. The peak observed at 708.50 eV corresponds to FeO ($Fe^{2+}$), whereas peak at 712.00 eV corresponds to a satellite transition that is characteristic feature of Fe *XPS* spectra [15, 16]. The presence of pure Fe metal peak is in accordance with earlier studies on LaFeAsO [17], $CaFe_2As_2$ [18], LaFePO [19]; however no satellite peak (712.00 eV) was observed in these studies. The absence of satellite peak (712.00 eV) and resemblance of 706.65 eV Fe metal peak was correlated with an itinerant character of Fe *3d* electrons [18]. However, the presence of satellite peak with $Fe^{2+}$ peak in studied samples and simultaneous appearance of Fe metal peak (706.65 eV) may be caused by charge-carrier localization induced by excess Fe at 2c site [12]. The existence of itinerant character of Fe *3d* electronic state caused by hybridization of *3d* states of Fe ions at 2a site and Se/Te 4p/5p states [18].

## CONCLUSIONS

In conclusion, we found that in $FeSe_{1/2}Te_{1/2}$ about 8% per mole of Fe occupies the interstitial 2c site. These Fe ions have localized magnetic moments which lead to a broad cusp like anomaly in electronic specific heat rather than well defined sharp lambda transition. This observation is further supported by our *XPS* measurements that show that Fe ions have two type of behavior. One, arising due to itinerant nature of Fe *3d* electrons due to hybridization of Fe *3d* and Se/Te 4p/5p states. Another, results from charge-carrier localization induced by excess Fe at 2c site.

### ACKNOWLEDGEMENTS

Anand Pal, A. Vajpayee and Jagdish Kumar thank CSIR (council of Scientific and Industrial research) India for Senior Research Fellowship. Bhasker Gahtori is supported by DST (Department of Science and Technology) FTP (Fast track position) for young scientists.

## FIGURE CAPTIONS

Figure 1: Rietveld fitted room temperature XRD pattern of $FeSe_{1/2}Te_{1/2}$, inset 1 shows the temperature dependence of electrical resistivity and inset 2 shows susceptibility vs. temperature of the same sample

Figure 2: Heat capacity behavior with temperature for $FeSe_{1/2}Te_{1/2}$. Upper inset gives entropy contribution due to superconducting condensate and lower inset shows observed $C_p$ data along with fitted normal state behavior up to 2 K.

Figure 3 (a): Fe 2p XPS spectra for $FeSe_{1/2}Te_{1/2}$. The dashed line represents experimental curve and red spheres represents resultant of fitted curves.

Figure 3 (b): Te 3d XPS spectra for $FeSe_{1/2}Te_{1/2}$. The dashed line represents experimental curve and red spheres represents resultant of fitted curves.

Figure 3 (c): Se 3d XPS spectra for $FeSe_{1/2}Te_{1/2}$. The dashed line represents experimental curve and red spheres represents resultant of fitted curves.

Figure 1: Rietveld fitted room temperature XRD pattern of $FeSe_{1/2}Te_{1/2}$, inset 1 shows the temperature dependence of electrical resistivity and inset 2 shows susceptibility vs. temperature of the same sample

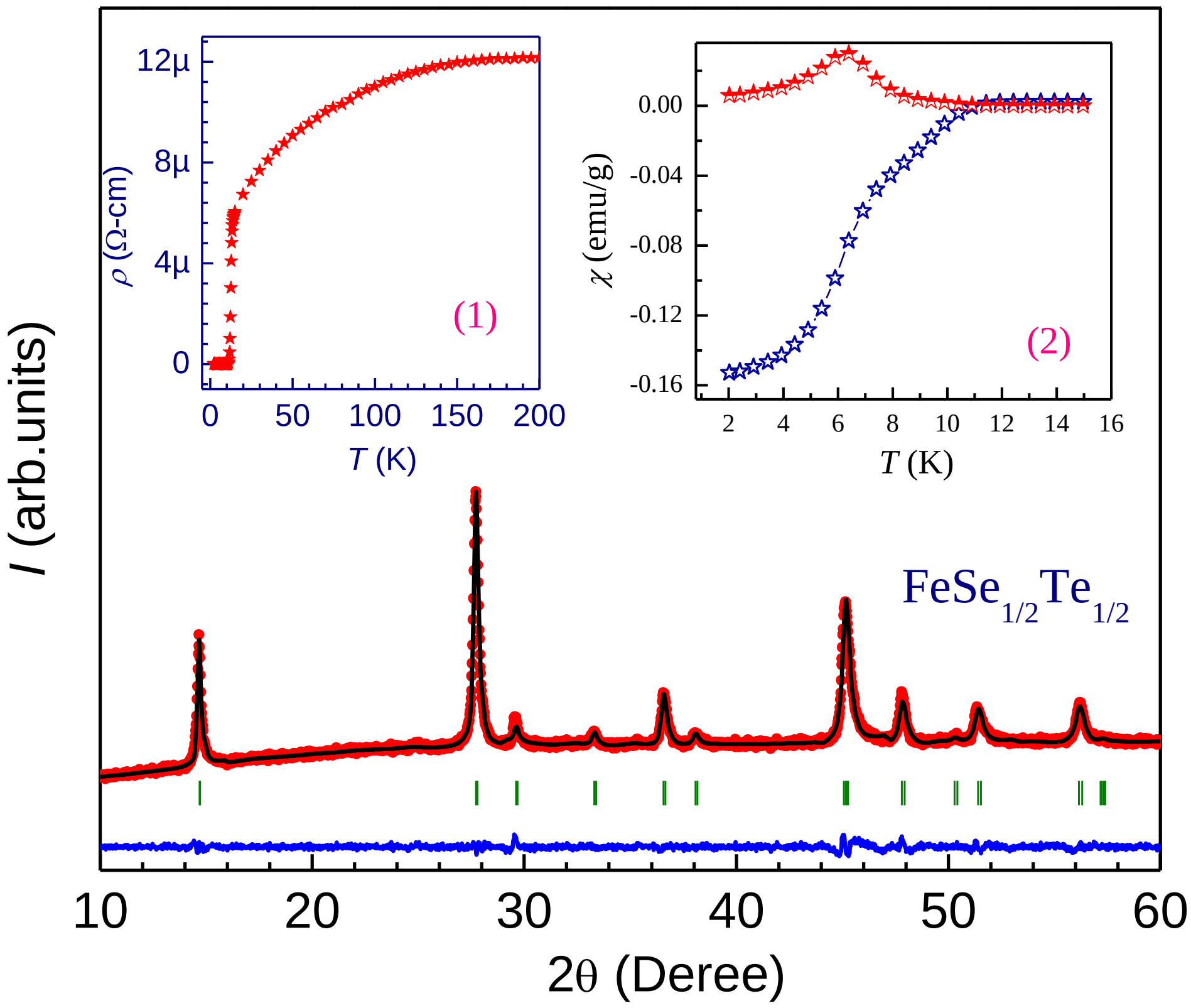

Figure 2: Heat capacity behavior with temperature for $FeSe_{1/2}Te_{1/2}$. Upper inset gives entropy contribution due to superconducting condensate and lower inset shows observed $C_p$ data along with fitted normal state behavior up to 2 K.

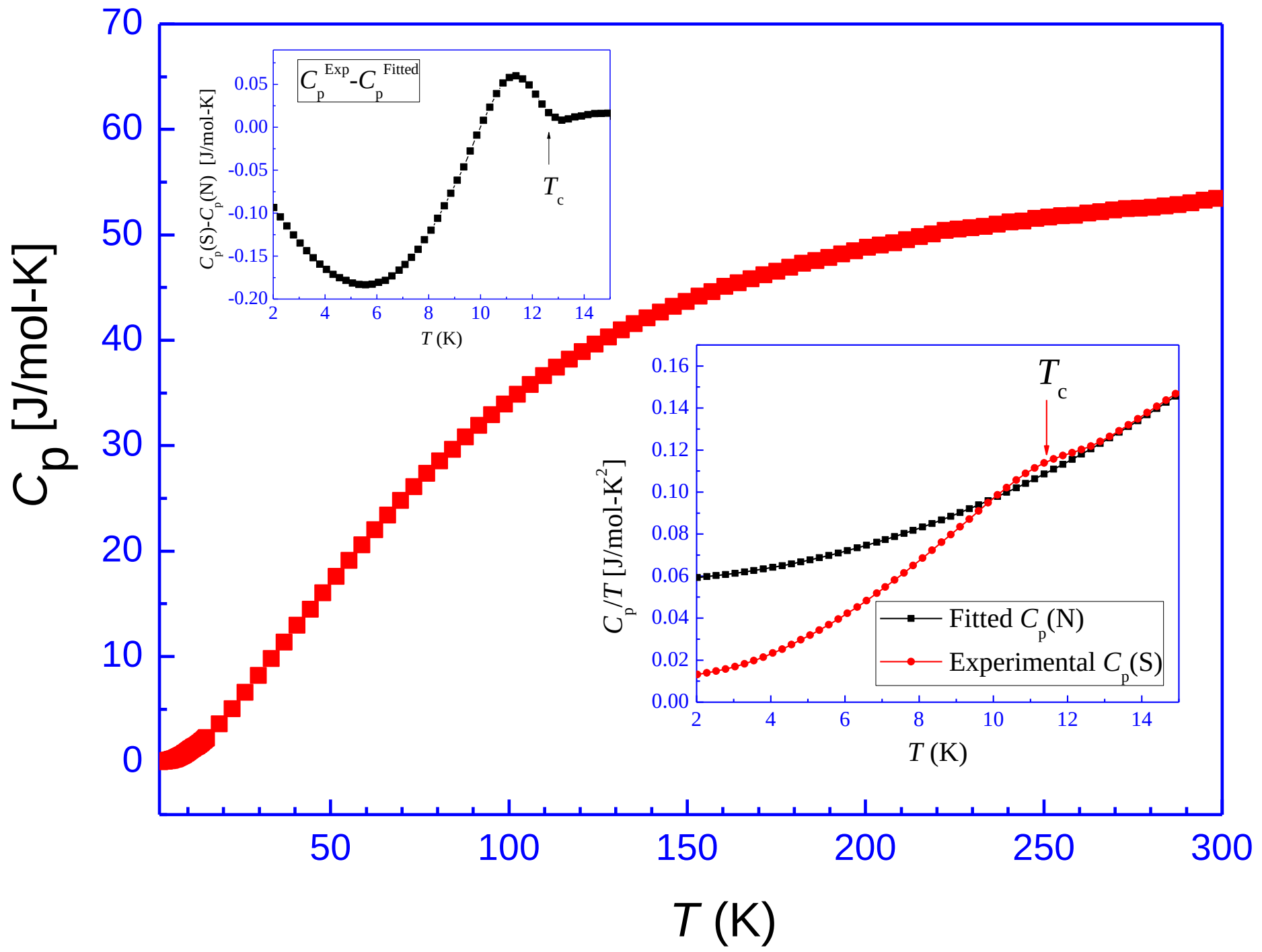

Figure 3 (a): Fe 2p XPS spectra for $FeSe_{1/2}Te_{1/2}$. The dashed line represents experimental curve and red spheres represents resultant of fitted curves.

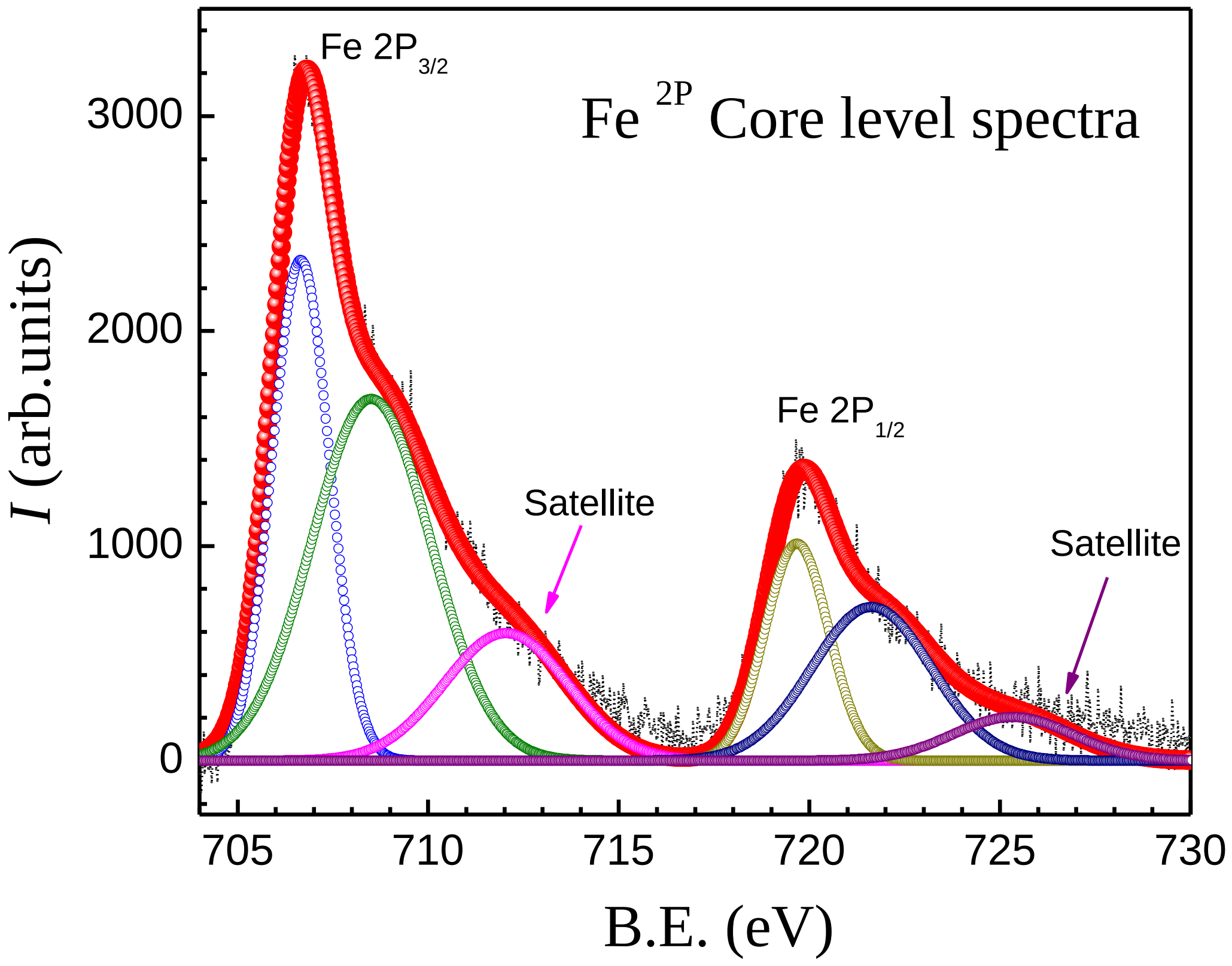

Figure 3 (b): Te 3d XPS spectra for $FeSe_{1/2}Te_{1/2}$. The dashed line represents experimental curve and red spheres represents resultant of fitted curves.

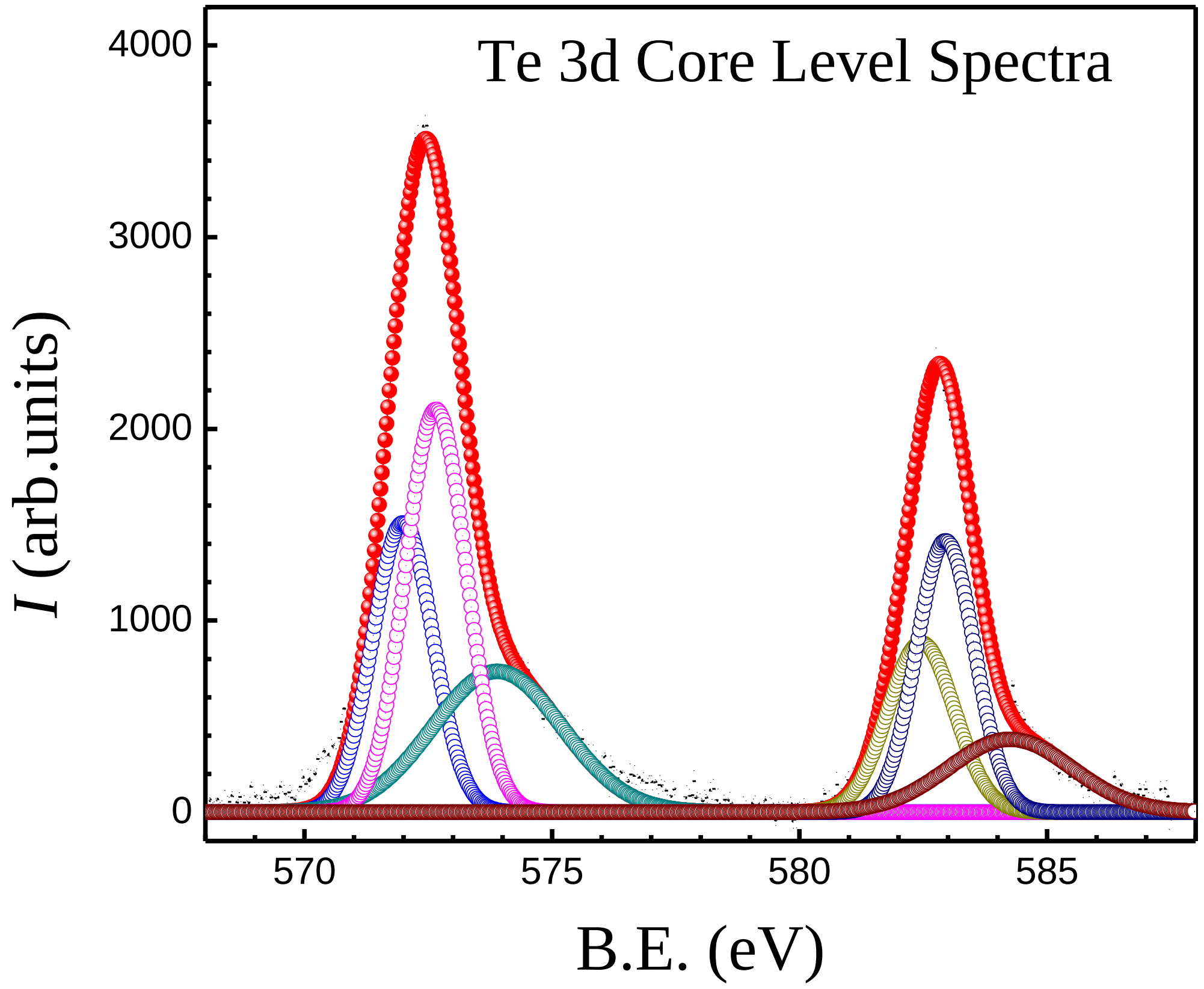

Figure 3 (c): Se 3d XPS spectra for $FeSe_{1/2}Te_{1/2}$. The dashed line represents experimental curve and red spheres represents resultant of fitted curves.

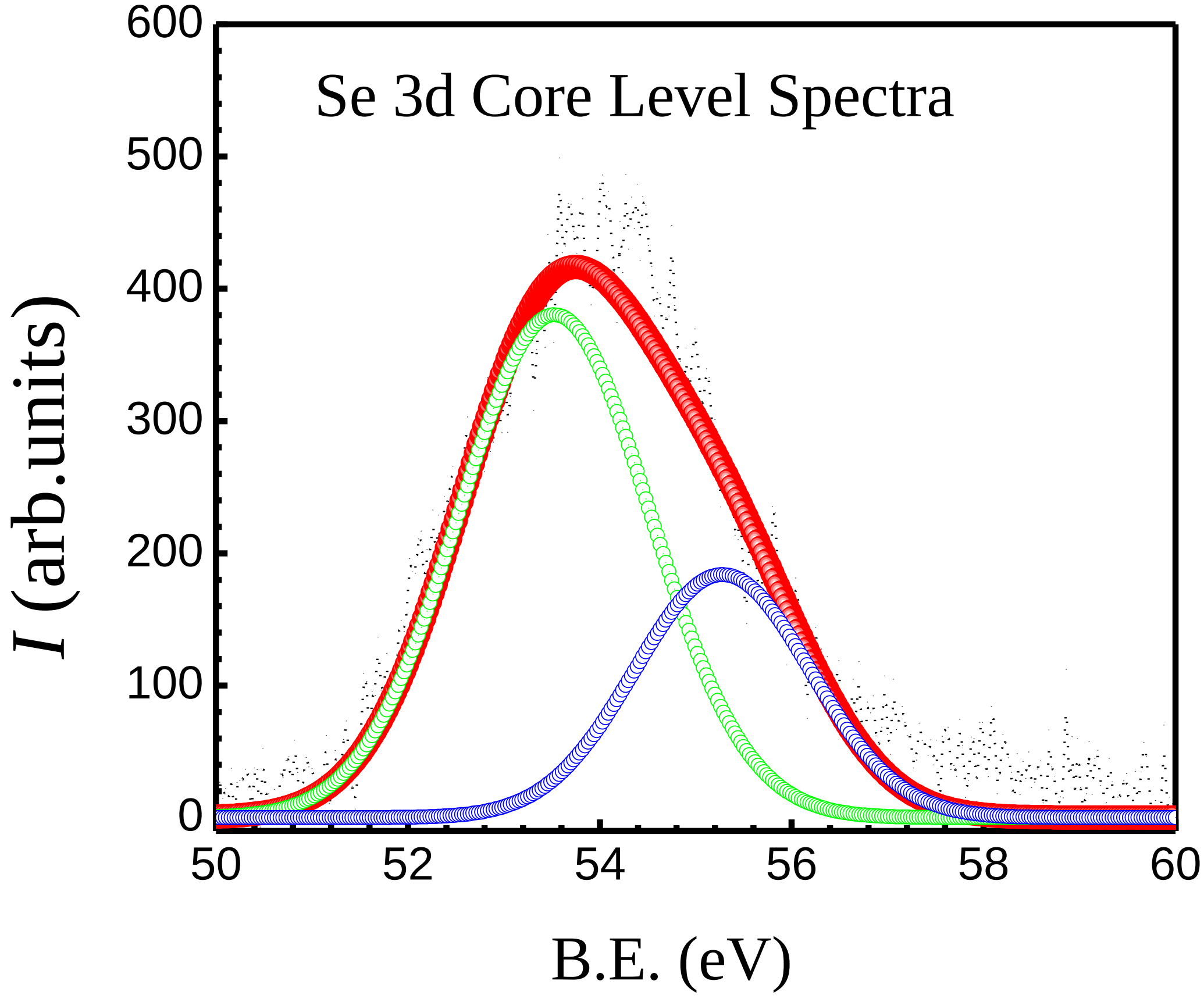